# Outdoor Field Experience with Autonomous RPC Based Stations

L. Lopes[b,*], P. Assis[a], A. Blanco[b], N. Carolino[b], M. A. Cerda[c,e], R. Conceição[a], O. Cunha[b], M. Ferreira[a], P. Fonte[b,d], R. Luz[a], L. Mendes[a], A. Pereira[b], M. Pimenta[a], R. Sarmento[f] and B. Tomé[a]

[a] *Laboratório de Instrumentação e Física Experimental de Partículas (Lip),*
   *Avenida Elias Garcia 14, 1º, 1000-149 Lisboa, Portugal*

[b] *Laboratório de Instrumentação e Física Experimental de Partículas (Lip),*
   *Departamento de Física da Universidade de Coimbra, 3004-516 Coimbra, Portugal*

[c] *Observatório Pierre Auger,*
   *Malargüe, Argentina*

[d] *Instituto Politécnico de Coimbra, ISEC*
   *Rua Pedro Nunes, 3030-199 Coimbra, Portugal*

[e] *Instituto Nazionale di Fisica Nucleare sezione di Roma Tor Vergata,*
   *Roma, Italy*

[f] *Laboratório de Instrumentação e Física Experimental de Partículas (Lip),*
   *Departamento de Física, Universidade do Minho, 4710-057 Braga, Portugal*

   *E-mail*: `luisalberto@coimbra.lip.pt`

ABSTRACT: In the last two decades Resistive Plate Chambers were employed in the Cosmic Ray Experiments COVER-PLASTEX and ARGO/YBJ. In both experiments the detectors were housed indoors, likely owing to gas distribution requirements and the need to control environment variables that directly affect RPCs operational stability. But in experiments where Extended Air Shower (EAS) sampling is necessary, large area arrays composed by dispersed stations are deployed, rendering this kind of approach impossible. In this situation, it would be mandatory to have detectors that could be deployed in small standalone stations, with very rare opportunities for maintenance, and with good resilience to environmental conditions. Aiming to meet these requirements, we started some years ago the development of RPCs for Autonomous Stations. The results from indoor tests and measurements were very promising, both concerning performance and stability under very low gas flow rate, which is the main requirement for Autonomous Stations. In this work we update the indoor results and show the first ones concerning outdoor stable operation. In particular, a dynamic adjustment of the high voltage is applied to keep gas gain constant.

---

[*] Corresponding author.

# Contents

## 1. Introdution

In the framework of the Pierre Auger Observatory upgrade, RPCs have been proposed as a dedicated muon detector to better estimate the muonic component of Extensive Air Showers (EAS), further constraining the nature of the cosmic rays and hadronic interactions taking place in the EAS development. An engineering array will be installed in the infill region in the next two years. Although this array will only allow the collection of a limited sample of moderate energy cosmic ray showers, it will be of extreme importance to set a calibration point as it allows a direct measurement of the muonic component of the showers. Furthermore, the Engineering array will be installed in the same region as AMIGA, an underground muon detector, allowing to cross-calibrate the two detectors. The instrumentation of 8 tanks as described in [8], will allow to study and improve the performance of Resistive Plate Chambers (RPCs) in outdoor inhospitable environment.

Resistive Plate Chambers [2] can be found in a large number of experiments, mainly in indoor but also in outdoor environments [3-6]. Five years ago [1, 8] we started research on the development of a chamber to operate outdoors with residual maintenance, low power and gas consumption and we believe to be now close to a robust and mature detector solution.

In this work we present the most important inputs and conclusions from indoor developments over the last 2 years and evaluate their applicability outdoors. The first outdoor results based on the dynamic adjustment of the high voltage will prove the possibility to operate these detectors at a "constant" efficiency, which is of main importance for this application.

## 2. Auger site environmental characterisation and detector module

The Pierre Auger Observatory is situated on the vast plain known as the *Pampa Amarilla* (yellow prairie) in the western Argentina [7], at an altitude of 1400 m above sea level. The climate is dry and relatively cold, the mean absolute pressure around 850 mbar, annual temperature oscillate between -20 °C in the winter and 35 °C in the summer. Wild daily temperature excursions, strong winds and thunderstorms occur very often.

Besides, deployment in remote field conditions requires that a special attention is given to mechanical robustness, ease of installation, simplicity of the gas system, low power and gas consumption and dense monitoring of operational and environmental variables.

The detector module consists of two 1 mm gas gaps defined by $1200 \times 1500 \times 1.9$ mm$^3$ glass electrodes separated by Nylon monofilaments. The stack is then closed inside a permanently glued acrylic box. The high voltage is applied by means of a layer of resistive acrylic paint [7] on the outer electrodes. Signals are read out through an 8x8 pad matrix, each pad with an area of $180 \times 140$ mm$^2$. The DAQ electronics, named PREC[14], is a custom-developed system based on discrete electronics. Each acquisition channel consists of a broadband amplifier followed by a programmable comparator. The threshold outputs go, via LVDS links, to a purely digital central board where data is buffered until read by the DAQ computer. A shielding aluminium box with volume 1650x1285x26 mm$^3$ gives mechanical structure and robustness to the module. A more detailed description of the detector module can be found in [8]. The expected plateau efficiency should be 90%, due to gap width and geometrical constrains [8]. High voltage (HV), background currents, temperature, pressure and relative humidity sensors are monitored each minute via an internal I$^2$C bus and the data recorded for analysis. Recent updates include the monitoring of the gas (pure R-134a) flow rates (optical and pressure-based systems for redundancy) and relative humidity at the outputs of the sensitive volume and aluminium box (figure 1). This way we can continuously check the tightness of both volumes and also confirm that we have the expected gas flow rate flushing the detector. Another important update already implemented is the housing of the frontend electronics and HV power supply inside the aluminium box.

## 3. Indoor characterization of the detector module under dynamic HV adjustment

Since the detector will operate outdoor is important to test and simulate as much as possible all outdoor expected situations. Cover temperatures from "zero" to 50 °C are easily reachable. Mimic pressure variations is more complicated, since a large external hypobaric chamber will be needed [9], which is not available at the moment. This way, indoor tests were always performed at absolute pressure around 1000 mbar, with the normal daily variations.

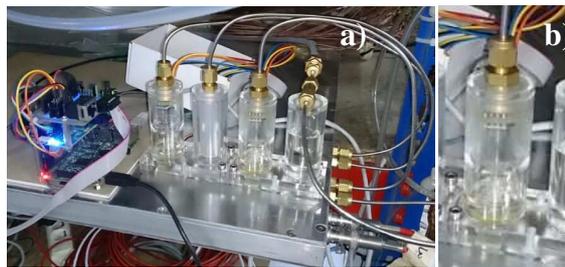

**Figure 1:** Picture of the gas monitoring unit. a) Columns from right to left: safety, output from sensitive volume, trap between sensitive volume output and aluminium box input, aluminium box output. b) Close view of the output column, the pressure and humidity sensors are on the upper part and the optical one is in the bottom inside the oil.

Other important issue to have in mind is the gas flow rate, if we are using calibrated holes directly connected to the bottle and not via a pressure regulator. The vapor pressure is temperature dependent and will directly define the pressure gradient over the calibrated micro hole. Since the dependence of the flow rate in this gradient is linear, this effect could become an important source of problems. This is even more important because we want to operate these detectors at a very low gas flow rate, in such a situation a factor 2 is of major importance, both with respect to chamber performance as with gas consumption. In spite of been a more "expensive" solution, the use of a pressure regulator and a capillary [8] instead of a calibrated hole should eliminate this risk.

All chambers are tested at very low gas flow rates at the end of the production line, measuring all practical quantities (except time resolution, which is far below the requested), as described in [8].

The chambers suffer the effect of daily and/or seasonal temperature and pressure variations, which renders impossible to operate them within the efficiency plateau. The analysis done in this section will show the importance of the dynamic adjustment of the HV to compensate the reduced applied electric field, E/N, from temperature and/or pressure variations. We will use the data from unit 23, which was used for a long run test before been shipped away. Normally validation tests take between 2 and 3 weeks depending on the preliminary Ar discharge cleaning and conditioning with Tetrafluorethane at 12 cc/min.

As in [8] the reduced electric field inside the gas gaps is given by:

$$\frac{E}{N} = 0.0138068748 \times \frac{V_{eff,Volts}}{d_{cm}} \frac{(T_{°C} + 273.15)}{P_{mbar}}, \quad [Td]$$

where $E$ is the applied electric field and $N$ is the gas numerical density. The unit Townsend (*Td*) is customary in the gaseous electronics and plasma fields.

Since the gap width, $d_{cm}$ is fixed, we only can use $V_{eff}$ to compensate from temperature and/or pressure variations. As known, $V_{eff}$ also depends on the potential drop across the resistive electrodes, which in turn also depends on temperature through the glass volume resistivity [13]. Since this is a low rate application and it is difficult to be sure that the total current comes only from the gaps we can neglect this effect for now and take $V_{eff} = V_{ap}$, this way [12],

$$V_{eff} = V_{ref} \times \frac{T_{ref}}{T_{measured}} \times \frac{P_{measured}}{P_{ref}}, \quad [V(V), T(°C), P(mbar)]$$

As known from [8], for *E/N = 238 Td* we are inside the efficiency plateau, considering $T_{ref}$ = *25 °C* and $P_{ref}$ = *1000 mbar* we get $V_{eff}$ = *5800 V/gap*, and the previous equation becomes:

$$V_{ap} = V_{eff} = 5800 \times \frac{25}{T_{measured}} \times \frac{P_{measured}}{1000}, \quad [V]$$

After some indoor and outdoor tests adjust the high voltage each 15 minutes, using the average data of the last 15 minutes, seems to be enough to accommodate pressure and temperature variations. In figure 2 we plot the data that support the considerations to reach the last equation; the arrow indicates the time when we start the automatic adjustment. As expected temperature variations define the long term and pressure variations the short term behaviour of the high voltage. As indicated in the figure, the test was performed for more than 9 months, most of the time at 1 cc/min without any evidence of abnormal behavior. For the variations observed on temperature and pressure, the correction on the gain only via the adjustment of the high voltage seems to be enough to get a stable E/N and consequently a stable gain. Other contributions to the gain are not considerer since their effect is too small within the considered variations in pressure and temperature. Since the gain is defined "only" by the reduced electric field, if it has a stable variation over time the same should be observed for the induced charge and efficiency. Figures 3 and 4 support these considerations.

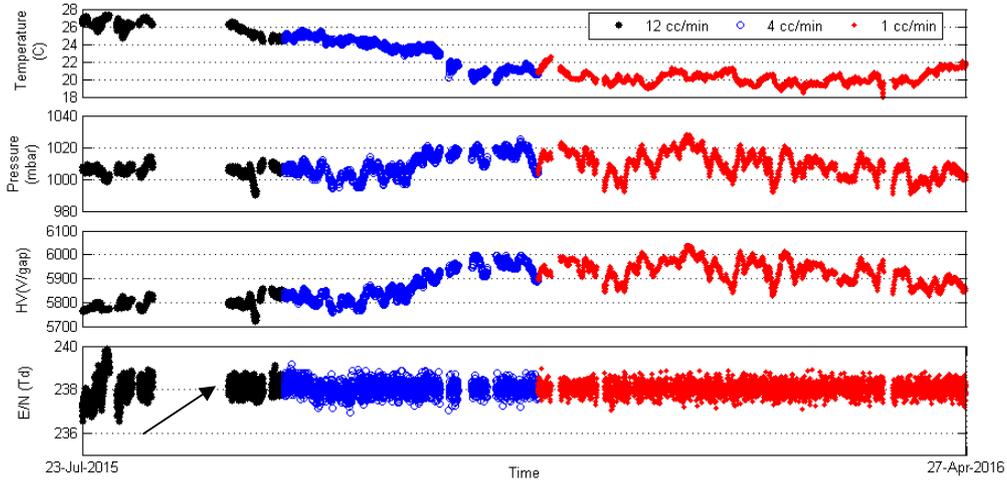

**Figure 2:** Reduced electric field and the three variables considered for its determination via the automatic adjustment of the applied high voltage. The arrow indicates the start of the adjustment process. The effect is clear over more than 8 months and as expected independence of the gas flow rate. Temperature defines the long term and pressure the short term behaviour of the applied high voltage.

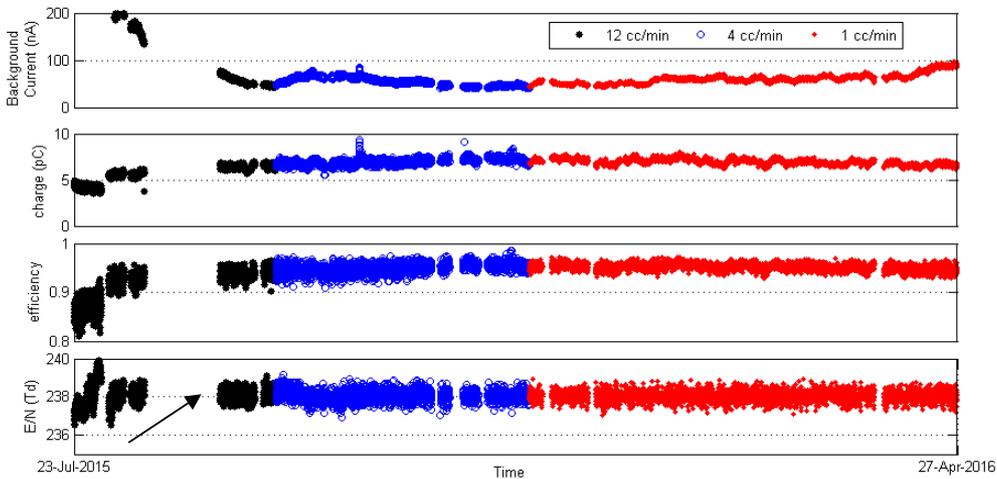

**Figure 3:** Reduced electric field, efficiency, fast charge and background (operation) current over more than 9 months. As expected, following the stability of the E/N we observe a stable fast charge and a stable efficiency. The current is not as stable as the other variables since it is the sum of various contributions; (see text)

In figure 3 we can observe the variation of the induced fast charge, efficiency, background current and E/N over more than 9 months. For small variations of temperature and pressure the dynamic adjustment of the high voltage allows for a stable reduced electric field, induced charge and efficiency over time. In figure 4 we show the distribution of these three variables

– 4 –

and observe narrow distributions for all of them. Hence we can expect to operate at a constant efficiency if we assure a narrow distribution of E/N through the dynamic adjustment of the high voltage.

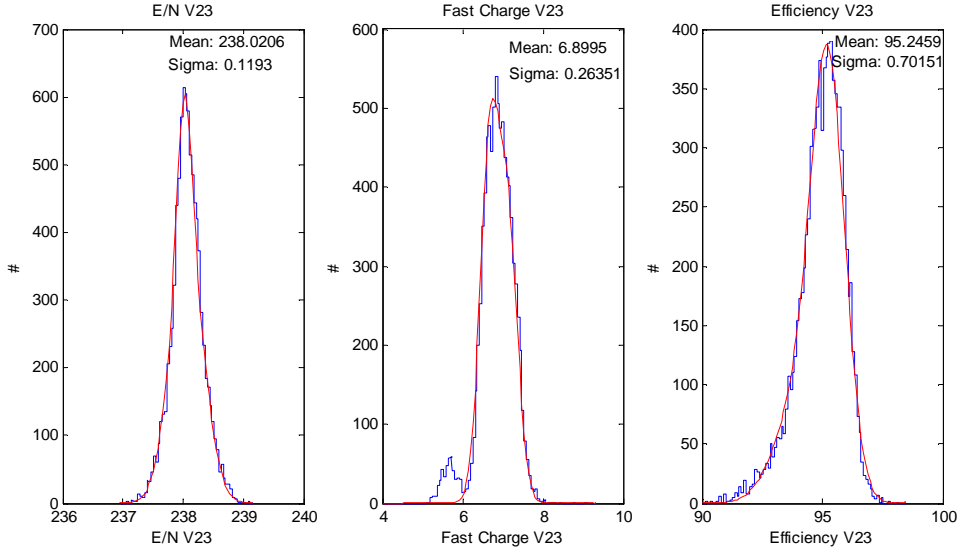

**Figure 4:** Distributions of the reduced electric field, fast charge and efficiency for the data taken over dynamic adjustment of the high voltage. Assuring a narrow distribution of E/N we can expect to operate with a constant efficiency, which is of major importance when we want counting particles over time.

The current in figure 3 is not as stable as the other variables since it is the sum of various contributions; ionization currents, leakage currents and also the current that comes from the increase of the background rate with the increase in temperature. The last two contributions will not affect the charge nor efficiency (analysis) since they are excluded by the trigger definition.

The impact of the temperature increase could become important for these wide gap chambers with very limited counting rate capability. For applications where temperature increases up to 50 ºC the background current (and rate) will increase too much and the consequent voltage drop in the resistive electrodes may be not totally compensated by the glass resistivity decrease due to the same increase in temperature. In a situation like that operate at a high and constant efficiency will be impossible, and one we would need to operate at a lower efficiency to ensure stability over time.

## 4. Field experience, first results

In January 2014 we installed the first two detectors in the Gianni Navarra (GN) tank at the AUGER Observatory [8]. The detectors were with HV on for most of the time thenceforth, just with two longer interruptions to repair HV leakages in the connectors due to condensation and minor interruptions due to problems with the gas supply. The aim of this setup is the study of the response of the water Cherenkov tank to single muons using an RPC hodoscope [10]. One detector is on the top and the other underneath the tank. The upper one is only protected from direct Sun and rain by a small roof; on the hottest days we can record temperatures inside the detector aluminum box above 40 ºC, and in the coldest days below -5 ºC, with daily excursions could exceed 25 ºC. The one underneath the tank is more protected and temperatures gradients are narrow, both to daily excursions and seasonal variations. The dynamic adjustment of the high voltage makes the operation at a stable efficiency feasible; however for the one in the top of the tank we need to consider some increase in the protection against Sun and wind or increase the frequency of adjustments, which is not as effective. The reason is related to the wide temperature daily excursions and thermal inertia of the detector; since the sensors are inside the Aluminium box any variations of the temperature are immediately recorded. However the thermal inertia of the sensitive volume slows the effect of these changes inside the gap. Despite all this observations the GN setup has been on (at 12 cc/min considering the operation conditions) for more than 2 years allowing the tank calibration and more important as a source of learning in the constant development of this RPC design. Figure 5 shows the variation over a week of the temperature, pressure, high voltage and coincidence rate between 9 pads of the top one and 6 pads of the bottom chamber in the setup reported in [10]. Despite wide temperature daily excursions the dynamic high voltage adjustment seems to be able to compensate the most part, leading to a nearly constant coincidence trigger rate.

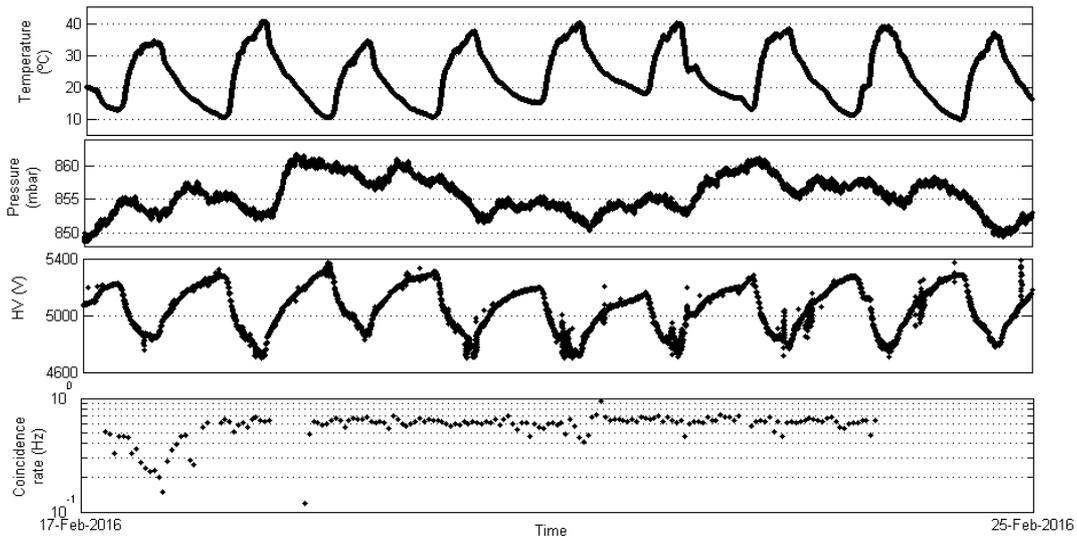

**Figure 5:** Variation over time of the temperature, pressure, high voltage and coincidence trigger rate between 9 pads of the top one and 6 pads of the bottom chamber at the GN hodoscope. Although wide temperature daily excursions the dynamic high voltage adjustment seems to be able to compensate the most part, leading to a nearly constant coincidence trigger rate

The first real installation in the field took place in April 2014 in the infill region of the Auger detector array. The name of the tank attributed to us, Tierra del Fuego (TdF), perfectly describes the inhospitable place. The detectors are housed inside a precast concrete structure underneath the tank. This precast structure also acts as filter to the electromagnetic component of the showers and supports the tank. This assembly is an important barrier against all environment variables except the relative humidity. We experimented some high voltage leakages as in the GN setup, but managed a solution "opening some windows" in the precast structure. Due the large thermal inertia surrounding the detectors some simulations suggested and the data confirms temperature daily excursions below 3 °C.
The precast is prepared to receive four RPCs covering the tank diameter, but since we want to measure the efficiency of the detectors to muons, we stacked one on the top of the other spaced by 10 mm. This way we can use the tank and one RPC to define the trigger and measure the efficiency of the other one. In figure 6 it can be observed the time dependence of the reduced electric field, applied high voltage, pressure and temperature on the left side and the reduced electric field (since March 2016) distribution on the right. This is for one of the detectors, but

– 6 –

both have the same behavior. The similarities to what we observed in the laboratory are evident, confirming the capability of the chosen assembly to assure a stable operation of detectors.

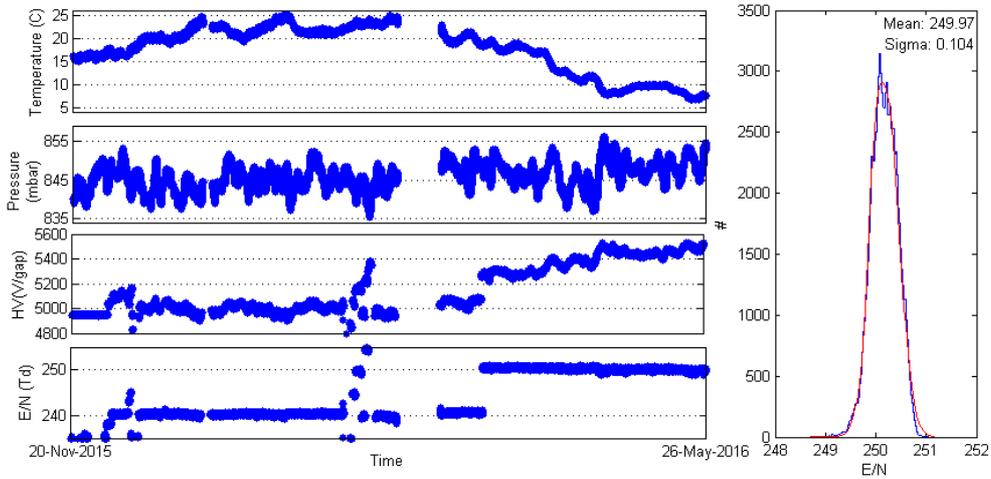

**Figure 6:** On the left side the reduced electric field and the three variables considered for its determination via the automatic adjustment of the applied high voltage. This detector operates at a gas flow rate of 4 cc/min. On the right side the distribution of the reduced electric field since March.

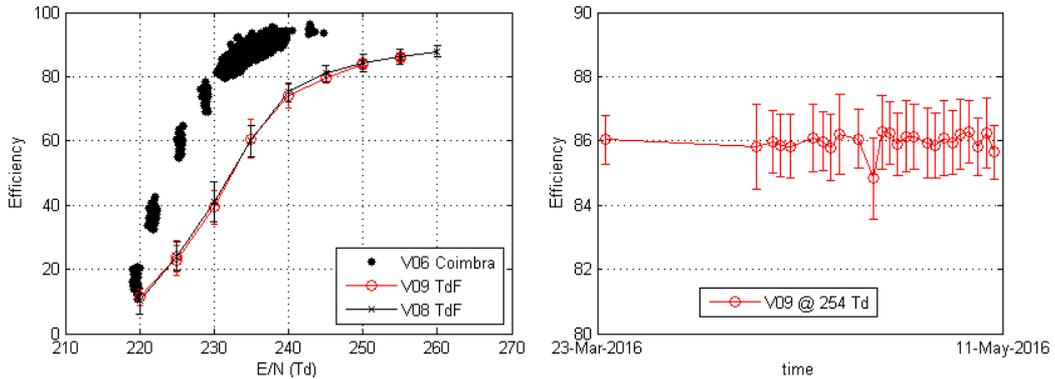

**Figure 7:** On the left we have the efficiency vs. E/N for the two detectors installed at Auger site and of one detector measured in the laboratory in Coimbra. To reach the same efficiency we need more 10 Td in the Auger site when compared to the Coimbra measurements. It is also clear that the efficiency plateau should be a few percent below. On the right we show the efficiency stability over time (point/day) for more than 1.5 months at a "constant" E/N of 254 Td.

When we started the dynamic high voltage adjustment at the end of November 2015 we have just taken as the reference E/N the one experimentally determined in the laboratory measurements, around 240 Td that should be within the efficiency plateau. However in the efficiency measurements (figure 7, left panel) it is observed a shift in E/N towards higher values to be able to reach the same efficiency as in the laboratory, more precisely 10 Td. Since we are operating at the same reduced electric field we would expect to get the same efficiency. In the laboratory [9] tested the performance at low pressures between 1000 and 400 mbar. They also use wide gaps and conclude that the efficiency plateau decreases with the pressure decrease, even when they equalize the effect of the pressure decreasing with the high voltage they were not able to superimpose all curves for the set of pressures under test. The explanation (considering the proportional avalanche limit) [11] lies in the fact that to keep everything constant under "large" gas density variations we need not only to adjust the high voltage to keep the gain constant as we do, but also correct the gap width to keep constant the ratio between the number of ionization clusters and the number of ionization steps. Only this way we assure the possibility to operate at the same efficiency plateau.

In the right panel of figure 7 we show the efficiency stability over time (point/day) for more than 1.5 months at a "constant" E/N of 254 Td. Like in the laboratory for normal daily variations of temperature and pressure and consequently of the gas density, if we keep E/N constant we will assure a constant efficiency, which is of main importance when we want to count particles over time.

## 5. Conclusion

In indoors conditions, the improvements made over the last two years have shown to be very important to deeply understand the operation of this detector design. A set of auxiliary systems were developed, to both control and monitor "all" variables that influence or could influence the detector performance. We prove the importance of the dynamic adjustment of the high voltage to ensure a constant reduced electric field over time. In this way we nearly fulfill one of the most important requirements for counting particles over time: a constant efficiency. Although within a short time period, the same observation was done outdoors in the GN and TdF setups.

Both outdoor setups are now running for more than two years without major problems. In the GN setup the detectors operate in very harsh conditions; extreme daily temperature excursions (over 25 ºC), strong winds that hit the chamber with all kind of dust and humidity.

Relatively to the gap physics, we now understand and confirm previous observations as well, which was predicted by theory on the effect of pressure in the chamber efficiency. Compensating pressure variations only with high voltage will not be enough, is also need to adjust the gap size to compensate the dependence of the ionization density on the pressure. The same is also valid for temperature variations.

**Acknowledgments**

This work is supported by Portuguese national funds OE, FCT-Portugal, CERN/FIS-NUC/0038/2015. The authors; P. Assis thanks FCT-Portugal for the grant SFRH/BPD/102031/2014; R. Conceição thanks FCT-Portugal for the grant SFRH/BPD/73270/2000; R. Luz thanks the FCT-Portugal and IDPASC Portugal for the grant PD/BD/113488/2015; R. Sarmento thanks the FCT-Portugal for the grant SFRH/BPD/84304/2012.